\documentclass{jkas}

\def\year{2015} 
\def\month{February} 
\def\volume{48} 
\def\issue{1} 
\def\beginpage{1} 
\def\endpage{12} 
\def\received{October 27, 2014} 
\def\accepted{November 27, 2014} 
\date{Received \received; accepted \accepted}

\def\ie{{i.e.,\ }}
\def\kms{~{\rm km~s^{-1}}}
\def\cm3{~{\rm cm^{-3}}}
\def\yr{~{\rm yr}}
\def\muG{~{\mu\rm G}}

\usepackage{flushend}

\title{Nonthermal Radiation from Relativistic Electrons\\ Accelerated at Spherically Expanding Shocks}

\author{Hyesung Kang}

\affil{Department of Earth Sciences, Pusan National University, Pusan 609-735, Korea; \email{hskang@pusan.ac.kr}}

\def\corrauthor{H. Kang}
\def\runningauthor{Kang}
\def\runningtitle{Radiation from Relativistic Electrons Accelerated at Spherically Expanding Shocks}
\def\keywords{acceleration of particles --- cosmic rays ---
galaxies: clusters: general --- shock waves}

\def\abstracttext{
We study the evolution of the energy spectrum of cosmic-ray electrons accelerated 
at spherically expanding shocks with low Mach numbers and the ensuing spectral 
signatures imprinted in radio synchrotron emission. Time-dependent simulations 
of diffusive shock acceleration (DSA) of electrons in the test-particle limit have 
been performed for spherical shocks with parameters relevant for typical 
shocks in the intracluster medium. 
The electron and radiation spectra at the shock location can be described properly 
by the test-particle DSA predictions with instantaneous shock parameters. 
However, the volume integrated spectra of both electrons and radiation 
deviate significantly from the test-particle power-laws, 
because the shock compression ratio and the flux of injected electrons 
at the shock gradually decrease as the shock slows down in time.
So one needs to be cautious about interpreting observed radio spectra of 
evolving shocks based on simple DSA models in the test-particle regime. 
}

\begin{document}
\jkashead 
\section{Introduction}

X-ray emitting baryonic gas in galaxy clusters is thought to be heated via
collisioness shocks that are induced during the formation of the large scale 
structure in the Universe \citep[e.g.,][]{ryu03,kang07,vazza09,skill11}.
The gas infalling toward nonlinear structures go through strong accretion shocks 
first, and later may encounter much weaker shocks produced by mergers and
chaotic flow motions inside the hot intracluster medium (ICM) \citep{ryu03}.
In fact, about a dozen of these weak ICM shocks have been detected through X-ray observations 
as bow shocks associated with merging activities
\citep[e.g., the so-called bullet cluster 1E 0657–56, ][]{markevitch02,akamatsu13,ob13}.

In addition, quiet a few ICM shocks have been identified as ``radio relic shocks'' 
in the outskirts of galaxy clusters through radio synchrotron radiation from cosmic-ray (CR)
electrons that are thought to be accelerated via diffusive shock acceleration (DSA) \citep{vanweeren10, vanweeren11, nuza12}.
Radio relics are interpreted as synchrotron emitting structures that contain relativistic electrons with
Lorentz factor $\gamma_e\sim 10^4$ gyrating in mean magnetic fields $B\sim 5-7 \muG$ in the ICM
\citep{kang12,feretti12, pinzke13, brunetti2014}.
The electrons are expected to be accelerated at weak shocks with the sonic Mach number, $M_s \sim 2-4$,
inferred from the spectral index $\alpha_{\rm inj} > 0.6$ of synchrotron emission immediately behind the shock.
Note that the spectral index of the volume-integrated synchrotron radiation is normally 
$A_{\nu}=\alpha_{\rm inj}+0.5$, since
electrons cool via synchrotron and inverse-Compton (iC) losses behind the shock \citep[e.g.,][]{kang11}.

It is well accepted that the DSA theory can explain how nonthermal particles are produced
through their interactions with MHD waves in the converging flows
across astrophysical shocks \citep{bell78, dru83, maldru01}.
In the {\it test particle limit} where the CR pressure is dynamically insignificant, 
it predicts that the CR energy spectrum at the shock
position has a power-law energy spectrum, $N(E) \propto E^{-s}$,
where $s=(\sigma+2)/(\sigma-1)$ and $\sigma=\rho_2/\rho_1$ is the shock compression ratio.
(Hereafter, we use the subscripts `1' and `2' to denote the
conditions upstream and downstream of shock, respectively.)
Hence, the DSA model can provide a simple and natural explanation for the spatially resolve
synchrotron radiation spectrum at the shock location, $j_{\nu}(x_s)\propto \nu^{-\alpha_{\rm inj}}$, 
where $\alpha_{\rm inj} = (s-1)/2$.

On the other hand, pre-acceleration of thermal electrons to suprathermal energies and the subsequent injection 
into the DSA process still remains very uncertain, especially at low Mach number shocks
in high beta ($\beta=P_g/P_B$) plasmas \citep[e.g.,][]{kang14}.
Recently, there have been some serious efforts to understand complex plasma instabilities and
wave-particle interactions operating at nonrelativistic shocks with the parameters relevant for weak cluster shocks,
through Particle-in-Cell (PIC) and hybrid plasma simulations \citep[e.g.,][]{riqu11,capri14a,capri14b,guo14}.
They have demonstrated that self-excitation of MHD/plasma waves via various instabilities, 
such as Bell's resonant and nonresonant instabilities and firehose instability, are crucial in the injection process.
Moreover, it has been shown that protons can be injected efficiently at quasi-parallel shocks, while
the electron injection occurs preferentially at quasi-perpendicular shocks.
Here, we adopt a {\it phenomenological} injection model in which suprathermal electrons with momentum
$p>p_{\rm inj}$ are allowed to cross the shock transition and participate in the Fermi 1st-order acceleration,
where the injection momentum, $p_{\rm inj}$, is several times the peak momentum of the postshock thermal protons
\citep{kjg02}.
This subject will be discussed in detail in Section 2.2.

The morphology of observed radio relics varies rather widely from diffuse patches to well-defined,
thin elongated structures.
For example, the so-called Sausage relic in CIZA J2242.8+5301 is located at a distance of $\sim 1.5$ Mpc 
from the cluster center and has an arc-like shape of $\sim 55$~kpc in width and $\sim 1-2$ Mpc in length \citep{vanweeren10}. 
So it would be natural to assume, at least for cases similar to the Sausage relic,
that radio relics consist of some portions of a spherical shell of radiating electrons
projected onto the sky plane.
The physical width of this shell is mainly determined by the cooling length of electrons, $\Delta l \sim  u_2 t_{\rm rad}(\gamma_e)$,
where $u_2$ is the postshock flow speed and $t_{\rm rad}$ is the electron cooling time.
But the projection of a partial shell could involve some geometrical 
complexities such as the extent of the partial shell and the viewing angle 
\citep[e.g.,][]{vanweeren10, vanweeren11,kang12}.

For several observed radio relics
the {\it projected} spatial profiles of the radio flux, $S_{\nu}$, and the spectral index, $\alpha_{\nu}$,
have been explained by the electron energy spectrum $N_e(r,\gamma_e)$ cooling radiatively
behind one-dimensional planar shocks \citep[e.g.,][]{vanweeren10,vanweeren11,kang12}. 
This may be justified, since the width ($\sim 50-100$~kpc) of those relics is much smaller than
their length ($\sim 1-2$ Mpc).
Note that a simple projection of a spherical shell on the sky plane was attempted in these previous papers,
although the DSA simulation results for a planar shock was adopted for $N_e(r,\gamma_e)$.

In this work we consider electron acceleration at spherical shocks with $M_s\sim 2.5-4.5$,
which expand into a hot uniform ICM.
A self-similar solution for Sedov-Taylor blast wave is adopted 
as the initial states of the {\it postshock} gasdynamic quantities (\ie $\rho$, $u$, $P$)
at the beginning of the simulations.
Later on the shock structure deviates from the Sedov-Taylor similarity solution, 
because the sonic Mach numbers of our model shocks are not large enough.
Nevertheless, the shock slows down approximately as $u_s(t) \propto t^{-3/5}$. 
This model may represent a spherically decelerating shock, but it is not meant to mimic
realistic cluster shocks.
Note that the flow speed, $u(r,t)$, is the only dynamical information that is fed into
the diffusion convection equation for the evolution of the electron distribution function, $f_e(r,p,t)$.
As the shock slows down and weakens, the shock compression ratio 
and the injection flux of CR electrons should decrease in time.
Consequently, the amplitude of electron spectrum at the injection momentum decreases
and the electron energy spectrum steepens gradually.

Considering that the synchrotron/iC cooling time scale ($t_{\rm rad}\sim 10^8 \yr$) for the electrons of 
our interest ($\gamma_e\sim 10^4$)
is shorter than the dynamical time scale of typical cluster shocks ($t_{\rm dyn} \sim 10^9 \yr$),
the electron energy spectrum at the shock location is expected to follow 
the test-particle DSA predictions with the {\it instantaneous} shock parameters.
But the spatial profile of the electron distribution function, $f_e(r,p)$,
and the volume-integrated distribution function, $F_e(p)=\int 4\pi f_e(r,p)~r^2 dr$,
would be affected by the time-dependent evolution of the spherically decelerating shock.
Our primary objective here is to study any signatures of such time-dependence through DSA simulations
of a heuristic example (\ie a spherical blast wave).
 
In the next section we describe the numerical method and
some basic physics of the DSA theory.
The simulation results of a planar shock and a spherical shock will be compared, 
and spherical shocks with different magnetic field models will be discussed in Section 3.
A brief summary will be given in Section 4.

\section{DSA Numerical Simulations}

\subsection{1D Spherical CRASH Code}

We consider DSA of CR electrons at nonrelativistic, gasdynamical shocks
in one-dimensional (1D) spherical geometry.
So we solve the following time-dependent diffusion-convection equation
for the pitch-angle-averaged phase space distribution function
for CR electrons, $g_e(r,p,t)=f_e(r,p,t) p^4$:
\begin{eqnarray}
{\partial g_e\over \partial t}  + u {\partial g_e \over \partial r}
= {1\over{3r^2}} {{\partial (r^2 u) }\over \partial r} ( {\partial g_e\over
\partial y} -4g_e)  \nonumber\\
+ {1 \over r^2}{\partial \over \partial r} [r^2 D(r,y)  
{\partial g_e \over \partial r}] 
+ p {\partial \over {\partial y}} \left( {b\over p^2} g_e \right),
\label{diffcon}
\end{eqnarray}
where $u(r,t)$ is the flow velocity, $y=\ln(p/m_e c)$, $m_e$ is the electron mass, $c$ is
the speed of light, and $D(r,p)$ is the spatial diffusion coefficient
\citep{skill75}. 
For $D(r,p)$, we adopt a Bohm-like diffusion coefficient with a weaker non-relativistic momentum
dependence
\begin{equation}
D(r,p) = 1.7\times 10^{19} {\rm cm^2s^{-1}} \left({ B(r)\over 1\muG}\right)^{-1} \left({p \over m_e c}\right).
\label{Bohm}
\end{equation}

The cooling term $b(p)=-dp/dt$ accounts for electron synchrotron and iC losses:
\begin{equation}
{1 \over t_{\rm rad} (\gamma_e)}= {b(p)\over p} =
\frac{4 e^4 }{9 m_e^3 c^5} B_{\rm e}^2  \cdot \gamma_e
\label{ecool}
\end{equation}
in cgs units, where $e$ is the electron charge.
The `effective' magnetic field strength $B_{\rm e}^2= B^2 + B_{\rm rad}^2$ takes account for 
radiative losses due to both synchrotron and iC processes,
where $B_{\rm rad}=3.24\muG(1+z)^2$ corresponds to the cosmic background
radiation (CBR) at redshift $z$  \citep{schlick02}. 
In this study, we set $z=0.2$ as a reference epoch and so $B_{\rm rad}=4.7\muG$.
Then the cooling time scale for electrons is given as
\begin{equation}
t_{\rm rad} (\gamma_e) = 3.8\times 10^{7} \yr \left({B_{\rm e} \over {8 \muG}}\right)^{-2} \left({\gamma_e \over 10^4 }\right)^{-1}.
\label{trad}
\end{equation}
For typical ICM conditions relevant for radio relics (\ie $n_H\sim 10^{-4} \cm3$ and $B\sim 1 \muG$),
the Coulomb loss dominates over the other losses for relativistic electrons with $\gamma_e \lesssim 100$,
while the synchrotron and iC losses are the main cooling processes for $\gamma_e  \gtrsim 100$ \citep{sarazin99}.
Nonthermal bremsstrahlung loss is much less important and can be ignored for relativistic electrons, 
since its cooling time scale is longer than the Hubble time \citep{petrosian01}.

Since we focus on the synchrotron emission from CR electrons here,
we do not need to consider the acceleration of CR protons.
Moreover, the CR proton pressure is expected to be dynamically insignificant 
at weak shocks with $M_s \lesssim 5$, so we can just follow the hydrodynamic evolution of
the shock without following DSA of CR protons.
This significantly alleviates the requirements for computational resources,
since much wider ranges of particle energy and diffusion length/time scales
should be included, if both proton and electron populations were to be calculated.
For instance, the maximum energy of electrons are limited to $\sim 100$ TeV
for the problem considered here, while that of protons can go up much higher energies
by several orders of magnitude.
  
At weak shocks in the test-particle limit, the CR feedback becomes negligible
and the background flow, $u(r,t)$, is governed by the usual gasdynamic conservations
in 1D spherical coordinates \citep{kj06}:
\begin{equation}
{\partial \rho \over \partial t}  +  {\partial\over \partial r} (\rho u) 
= -{ 2 \over r} \rho u,
\label{masscon}
\end{equation}
\begin{equation}
{\partial (\rho u) \over \partial t}  +  {\partial\over \partial r} (\rho u^2 + P) 
= -{2 \over r} \rho u^2,
\label{mocon}
\end{equation}
\begin{equation}
{\partial (\rho e_g) \over \partial t} + {\partial \over \partial r} 
(\rho e_g u + P u) =  -{2 \over r} (\rho e_g u + P u), 
\label{econ}
\end{equation}
where $P$ is the gas pressure, $e_{\rm g} = {P}/{[\rho(\gamma_{\rm g}-1)]}+ u^2/2$
is the total energy of the gas per unit mass and the rest of variables have
their usual meanings. The gas adiabatic index is assumed to be $\gamma_{\rm g}=5/3$,
since the background gas flow is nonrelativistic.

In order to optimize the shock tracking scheme, 
a comoving frame that expands with the instantaneous shock speed is adopted.
In such a frame, a spherically expanding shock 
can be made to be stationary by adopting comoving variables which factor out a uniform expansion. 
The details of the CRASH (Cosmic-Ray Amr SHock) code in a comoving spherical grid
can be found in \citet{kj06}.

\begin{figure*}[t!]
\centering
\includegraphics[trim=5mm 40mm 5mm 15mm, clip, width=135mm]{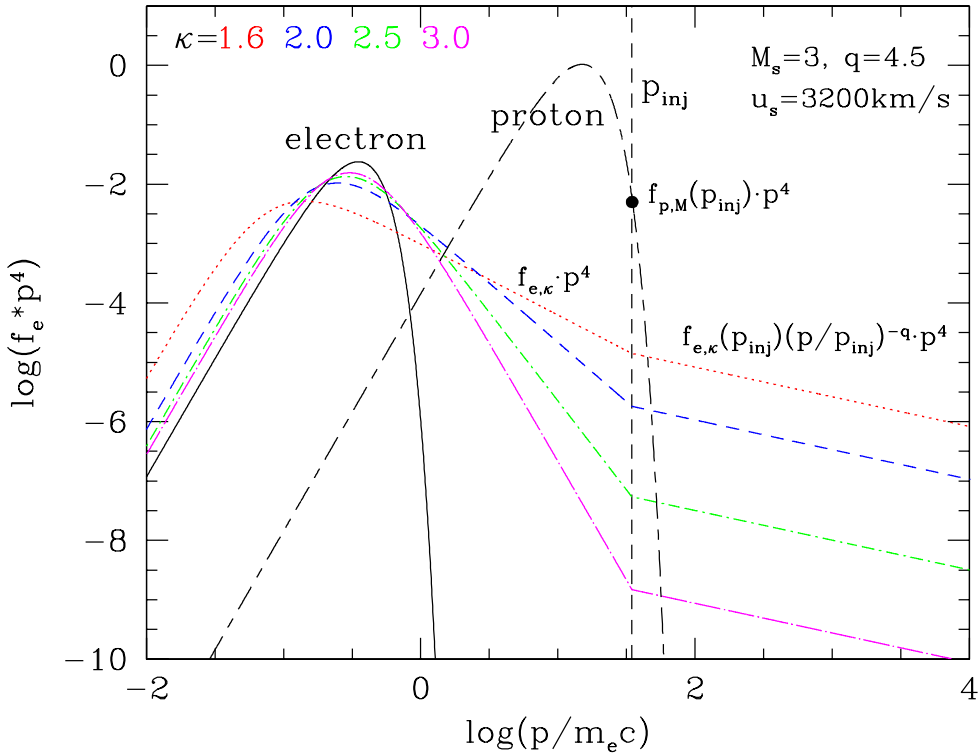}
\caption{ 
Schematic model demonstrating the pre-heated suprathermal distributions and the power-law distributions 
of electrons accelerated at a weak shock of $u_s=3.2\times10^3\kms$ and $M_s=3$.
The Maxwellian distribution (black solid line) and $\kappa$-distributions with
$\kappa_e=$ 1.6 (red dotted), 2.0 (blue dashed), 2.5 (green dot-dashed), and 3.0 (magenta dot-long dashed)
are shown for electrons.
For protons only the Maxwellian distribution (black dot-long dashed line) is shown.
The vertical line demarcates the injection momentum, so
$f_e(p)=f_{e,\kappa}(p)$ for $p<p_{\rm inj}$
and $f_e(p)=f_{e,\kappa}(p_{\rm inj}) \cdot (p/p_{\rm inj})^{-q}$ for $p\ge p_{\rm inj}$.
The filled circle marks the amplitude of $f_{p,M}(p_{\rm inj})$ at the injection boundary for the proton
Maxwellian distribution.
}
\end{figure*}


\subsection{Electron Pre-heating and Injection}

Injection of suprathermal particles into the Fermi process remains one of the outstanding problems in
the DSA theory.
In the so-called thermal leakage injection model, suprathermal particles with 
$p \gtrsim p_{\rm inj}\sim 3 p_{\rm th,p}$ are expected to re-cross the shock upstream
and participate in the DSA process,
because the shock thickness is of the order of the gyroradius of postshock thermal protons \citep[e.g.,][]{kjg02}.
Here, $p_{\rm th,p}=\sqrt{2m_p k_B T_2}$ is the most probable momentum of
thermal protons with postshock temperature $T_2$ and  $k_B$ is the Boltzmann constant.
On the other hand, recent hybrid and PIC simulations of collisioness shocks have revealed a somewhat
different picture, in which
incoming protons and electrons are specularly reflected at the shock surface and scattered
by plasma waves excited in the foreshock region, leading to suprathermal populations 
\citep[e.g.,][]{guo13, guo14,capriolietal14}.
The trajectories of these suprathermal particles are non-diffusive and confined to the region
mostly upstream of the shock transition.

If one assumes that downstream electrons and protons have Maxwellian distributions defined with 
the same kinetic temperature, $T_e\approx T_p$, then thermal electrons have much higher speeds but
much smaller momenta compared to thermal protons (\ie $p_{\rm th,e}=(m_e/m_p)^{1/2} p_{\rm th,p}$).
So electrons must be pre-accelerated from their thermal momenta
to $p_{\rm inj}\sim 3 p_{\rm th,p} \sim 130 p_{\rm th,e}$,
before they can begin to take part in the DSA process.
In fact, pre-heating of electrons above the thermal distribution have been widely observed in
space and laboratory plasmas \citep[e.g.,][]{vasyliunas68}.
Such suprathermal non-Maxwellian distributions can be described
by the combination of a Maxwellian-like core and a power-law tail,
which is known as the $\kappa$-distribution:
\begin{eqnarray}
f_{\kappa}(p)= {n_2 \over \pi^{1.5}}~ p_{\rm th}^{-3} { {\Gamma(\kappa+1) }
\over {(\kappa-3/2)^{3/2}\Gamma(\kappa-1/2) } }  \nonumber\\
\cdot \left[1+{p^2\over {(\kappa-3/2)p_{\rm th}^2}}\right]^{-(\kappa+1)},
\label{fkappa}
\end{eqnarray}
where $p_{\rm th}= \sqrt{2 m k_B T_2}$ is the thermal peak momentum,
the mass of the particle is $m=m_e$ for electrons and $m=m_p$ for protons,
and $\Gamma(x)$ is the Gamma function \citep[e.g.,][]{pierrard10}.
In the limit of large $\kappa$, it asymptotes to the Maxwellian distribution.

The power-law index of the $\kappa$-distributions for electrons and protons, 
$\kappa_e$ and $\kappa_p$, respectively, should depend on
plasma and shock parameters, \ie $\beta$, $M_s$, $M_A=u_s/v_A$ (where $v_A=B/\sqrt{4\pi \rho}$ is the Alfv\'en speed),
and the shock obliquity angle, $\Theta_{\rm Bn}$.
For example, the electron distributions measured near interplanetary shocks can be fitted with 
the $\kappa$-distributions with $\kappa_e\sim 2-5$, 
while the proton distributions prefer a somewhat larger $\kappa_p  \gtrsim 10$
\citep[e.g.,][]{pierrard10}.

As an illustration, Fig. 1 compares the electron $\kappa$-distributions, 
$f_{e,\kappa}(p)$ with $\kappa_e= 1.6-3.0$ for $p \le p_{\rm inj}$
and the electron power-law distribution, $f_{e,\kappa}(p_{\rm inj})\cdot (p/p_{\rm inj})^{-q}$ for $p > p_{\rm inj}$.
Here the shock parameters adopted are $u_s=3.2\times 10^3 \kms$, $M_s=3$, and the DSA power-law slope, $q=3$.
The Maxwellian distributions for electrons and protons are also plotted for comparison.
These $\kappa$-distributions may represent the pre-heated electron populations,
while the power-law distributions may represent the shock accelerated electron populations. 
We note that pre-heated electrons are assumed to be injected at $p=p_{\rm inj}$ with the amplitude, 
$f_{e,\kappa}(p_{\rm inj})$,
which is much larger than the corresponding value for the electron Maxwellian distribution, $f_{e,M}(p_{\rm inj})$.
Here we adopt $ p_{\rm inj} \approx 5.34 m_p u_2$, which becomes $ p_{\rm inj}\approx 3.3 p_{\rm th}$ for $M_s=3$.
The filled circle on the proton Maxwellian distribution marks the amplitude of proton distribution, 
$f_{p,M}(p_{\rm inj})$, at the injection
momentum, which is larger than $f_{e,\kappa}(p_{\rm inj})$ by several orders of magnitude.

For the case shown in Fig. 1, the ratio of CR electron to proton numbers can be approximated by
$K_{e/p}\approx {{f_{e,\kappa}(p_{\rm inj})}/{f_{p,M}(p_{\rm inj})}}$, 
which becomes $K_{e/p}\sim 1/300$ for $\kappa_e=1.6$.
So Fig. 1 demonstrates that the ratio becomes $K_{e/p}\sim 10^{-3}-10^{-2}$, depending on the value of $p_{\rm inj}$,
if the electrons are pre-accelerated to $\kappa$-distributions with $\kappa_e \lesssim 2$.

Since the main goal of this study 
is to explore how the electron energy spectrum and the synchrotron emission are
affected by the dynamical evolution of spherically expanding shocks (\ie $u_s(t)$ and $r_s(t)$),
the specific value of the amplitude $f_{e,\kappa}(p_{\rm inj})$
for a given set of models parameters (\ie $\kappa_e$ and $p_{\rm inj}/p_{\rm p, th}$) is not important.
So we simply assume that the injection momentum decreases with the shock speed
as $p_{\rm inj}(t) \approx 5.34 m_p u_s(t)/\sigma$, and the electrons are injected with
the value $K_{e/p}=1$ in the simulations described below.
Note that we do not attempt to compare the theoretically estimated radio flux, $S_{\nu}$,
with the observed radio flux of any specific radio relics.

\subsection{Shock Accelerated CR Electron Spectrum}

In the test-particle regime of DSA, the distribution of CR distribution function at the shock position
can be approximated, once it reaches equilibrium, by a power-law spectrum with super-exponential cutoff, 
\begin{equation}
f_{e,2}(p) \approx f_{\rm inj}\cdot \left(p \over p_{\rm inj} \right) ^{-q} \exp\left(-{p^2 \over p_{\rm eq}^2} \right), 
\label{f2o}
\end{equation}
where the power-law slope is $q = 3 \sigma /(\sigma -1 )$ and
$f_{\rm inj}$ is the amplitude at $p=p_{\rm inj}$ \citep{kang11}. 
The cutoff momentum can be derived from the equilibrium condition that the DSA momentum gains per cycle
are equal to the synchrotron/iC losses per cycle \citep{kang11}:
\begin{equation}
p_{\rm eq}= {m_e^2 c^2 u_s \over \sqrt{4e^3q/27}} \left({B_1 \over {B_{\rm e,1}^2 + B_{\rm e,2}^2}}\right)^{1/2}.
\label{peq}
\end{equation} 
The corresponding Lorentz factor for typical cluster shock parameters is then 
\begin{equation}
\gamma_{e, {\rm eq}} \approx {2\times 10^9 \over q^{1/2}} \left({u_s \over {3000 \kms}}\right) \left({B_1 \over {B_{\rm e,1}^2 + B_{\rm e,2}^2}}\right)^{1/2}.
\label{gammaeq}
\end{equation}

\begin{figure*}[t!]
\centering
\includegraphics[trim=3mm 5mm 5mm 5mm, clip, width=135mm]{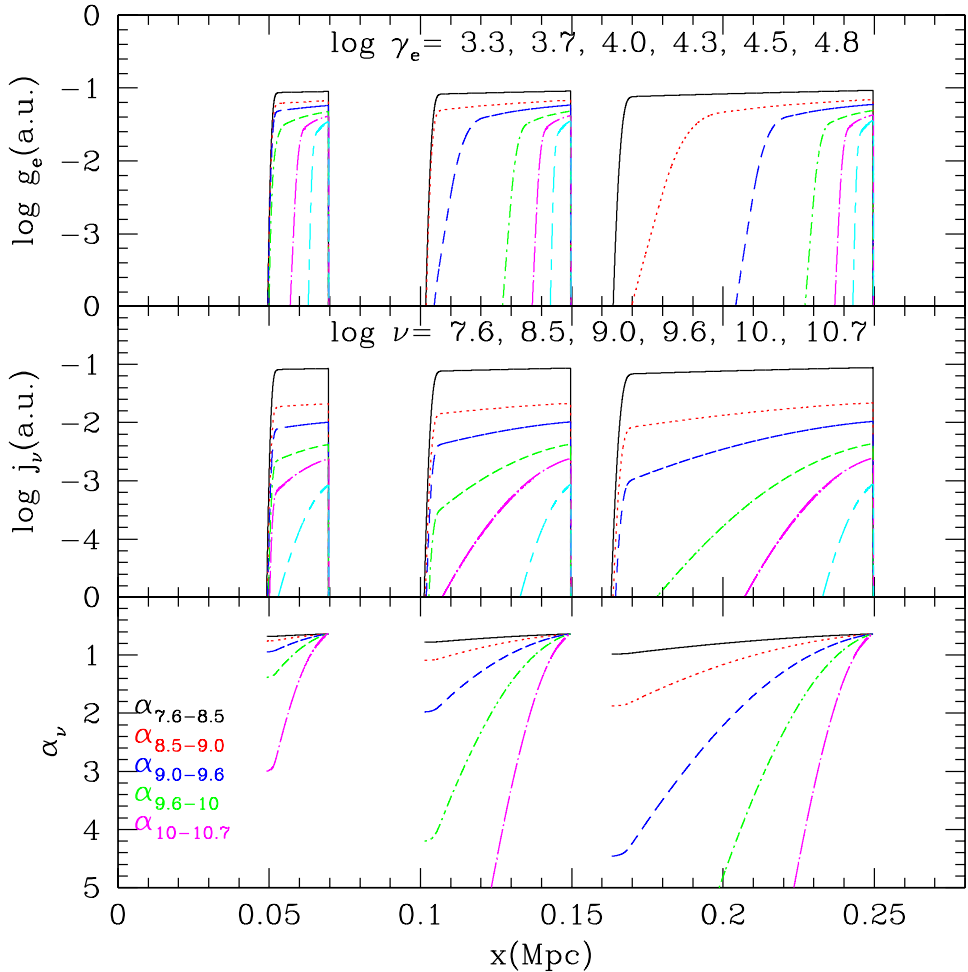}
\caption{ 
Time evolution of the planar shock of $u_s=3.5\times 10^3 \kms$, $M_s=4.5$, and $B_2=7\muG$
is shown for $t_{\rm age}= (1.8, 4.4, 8.0)\times 10^7 \yr$ (from left to right).
For clarity the shock location is shifted to the right by arbitrary amount at different time epochs.
{Top}:
Spatial distributions of the electron distribution function, $g_e(x,\gamma_e)$, for $\log(\gamma_e)=3.3$ (black solid lines), 3.7 (red dotted),
4.0 (blue dashed), 4.3 (green dot-dashed), 4.5 (magenta dot-long dashed), and 4.8 (cyan dash-long dashed).
{Middle}:
Spatial distributions of the synchrotron emissivity, $j_{\nu}(x,\nu)$ for $\log(\nu)=7.6$ (black solid lines), 8.5 (red dotted),
9.0 (blue dashed), 9.6 (green dot-dashed), 10. (magenta dot-long dashed), and 10.7 (cyan dash-long dashed).
{Bottom}: 
Spectral index $\alpha_{\log \nu_1 - \log \nu_2}$, where ($\log \nu_1$,$\log \nu_2$) is
(7.6,8.5) (black solid line), (8.5,9.0) (red dotted), (9.0,9.6) (blue dashed), (9.6,10.) (green dot-dashed), 
and (10.,10.7) (magenta dot-long dashed).
}
\end{figure*}

Postshock electrons cool radiatively while advecting downstream, so
the cutoff of the electron momentum spectrum decreases as one moves away from the shock.
As a result, the volume integrated electron energy spectrum, $F_e(p)$, 
becomes steeper than the power-law in Eq. (\ref{f2o}) by one power of the momentum 
above the `break momentum' ($p>p_{\rm e,br}$).
At the shock age $t$, the `break Lorentz factor' can be estimated from the condition 
$t_{\rm age}=t_{\rm rad} =p /b(p)$:
\begin{equation}
\gamma_{\rm e,br}(t)  \approx  10^4 \left({t_{\rm age} \over 10^8 \yr}\right)^{-1} \left({B_{\rm e,2} \over
{5 \muG}}\right)^{-2}.
\label{pbr}
\end{equation}

The synchrotron emission from mono-energetic electrons with $\gamma_{\rm e}$ peaks around
the frequency $\nu_{\rm peak} \approx 0.3 (3eB/4\pi{m_e c}) \gamma_e^2$, so it will be useful to
have the following relation:
\begin{equation}
\nu_{\rm peak}\approx 0.63{\rm GHz} 
 \left({ \gamma_e \over {10^4}}\right)^{2} \left({B \over 5\muG} \right).
\label{fpeak}
\end{equation}
Then the synchrotron radiation spectrum emitted by the electron population with 
the power-law given in Eq. (\ref{f2o}) also has a power-law form of
$j_{\nu}\propto \nu^{-\alpha_{\rm inj}}$ with $\alpha_{\rm inj} =(q-3)/2$.

For {\it diffuse} radio structures such as radio halos and radio relics
with $B\sim 5 \muG$ and the radio flux density, $S_{\nu} \sim 5~{\rm mJy}$,
the optical depth for synchrotron emission is small for the frequencies considered here ($\nu >10$~MHz)
\citep{jones74,lang99}.
So the synchrotron self-absorption can be ignored in the calculation of synchrotron radiation spectrum of the model shocks.

\begin{figure*}[t!]
\centering
\includegraphics[trim=3mm 5mm 5mm 5mm, clip, width=135mm]{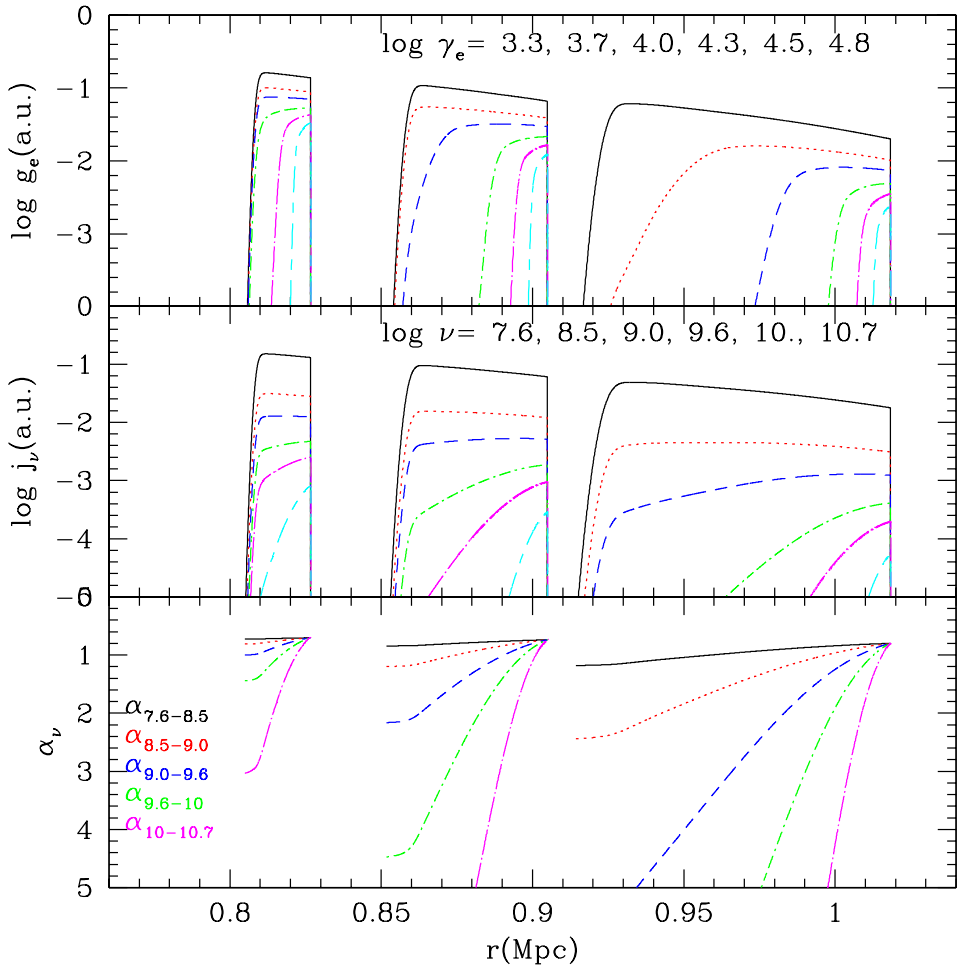}
\caption{ 
Time evolution of the spherical shock of $u_s=2.3\times 10^3 \kms (t/t_o)^{-3/5}$,
$M_s=3.0(t/t_o)^{-3/5}$, and $B_2=7\muG$ (where $t_o=1.73\times 10^8\yr$)
is shown for $t_{\rm age} = (1.7, 4.3, 8.7)\times 10^7 \yr$ (from left to right).
Electrons are injected at the shock from $t_i=0.5 t_o$ when $u_{s,i}=3.5\times 10^3\kms$.
{Top}:
Spatial distributions of $g_e(r,\gamma_e)$.
{Middle}:
Spatial distributions of $j_{\nu}(r,\nu)$.
{Bottom}: 
Spectral index $\alpha_{\log \nu_1 - \log \nu_2}$. 
The line types are the same as in Fig. 2.
}
\end{figure*}
\begin{figure*}[t!]
\centering
\includegraphics[trim=3mm 25mm 5mm 5mm, clip, width=135mm]{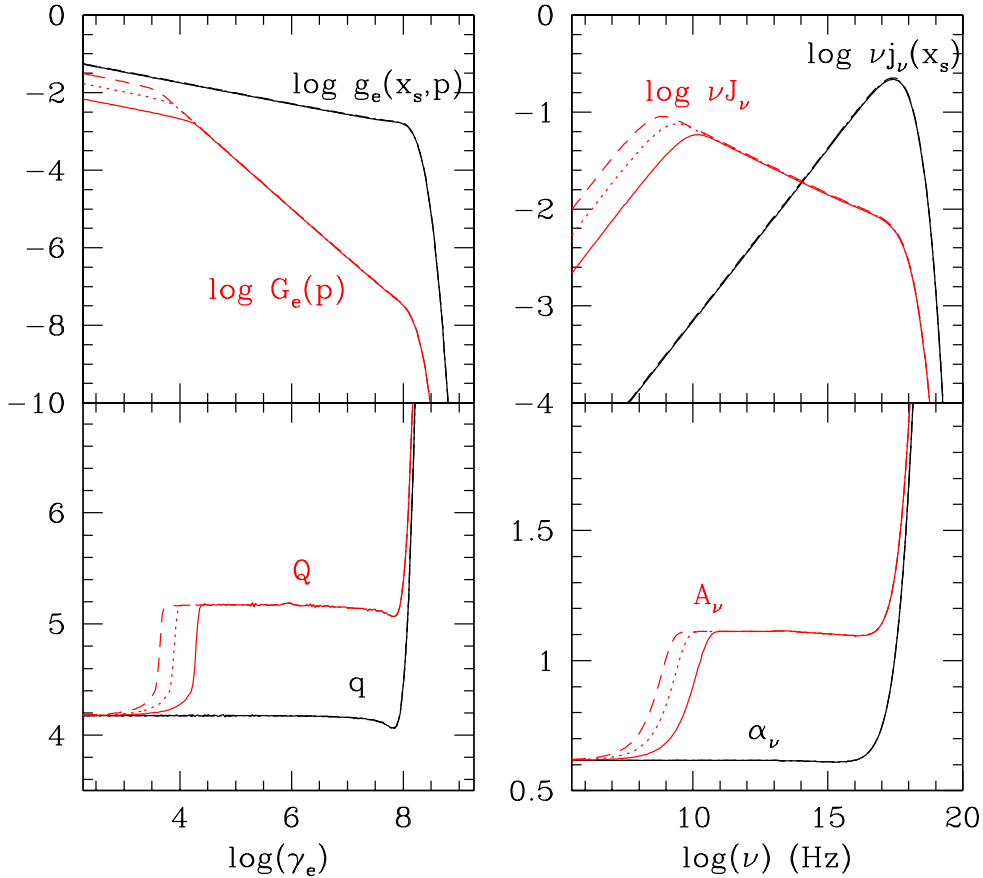}
\caption{ 
Electron (left) and synchrotron (right) spectral properties for the planar shock shown in Fig. 2
are plotted for $t_{\rm age}= (1.8, 4.4, 8.0)\times 10^7 \yr$ (solid, dotted, and dashed lines, respectively).
Upper left: 
electron distribution function at the shock position, $g_e(x_s,p)=p^4 f_e(x_s,p)$, 
and integrated over the simulation volume, $G_e(p)=p^4 F_e(p)$.
Lower left: 
electron distribution slopes, $q=-d \ln f_e(x_s)/d \ln p$ and
$Q=-d \ln F_e/d \ln p$.
Upper right: 
synchrotron radiation spectrum at the shock position, $\nu j_{\nu}(x_s)$,
and integrated over the simulation volume, $\nu J_{\nu}$.  
Lower right: synchrotron spectral indexes, $\alpha_{\nu}= d \ln j_{\nu}(x_s)/d \ln \nu $ 
and $A_{\nu}= -d \ln J_{\nu}/d \ln \nu $.
}
\end{figure*}

\begin{figure*}[t1]
\centering
\includegraphics[trim=3mm 25mm 5mm 5mm, clip, width=135mm]{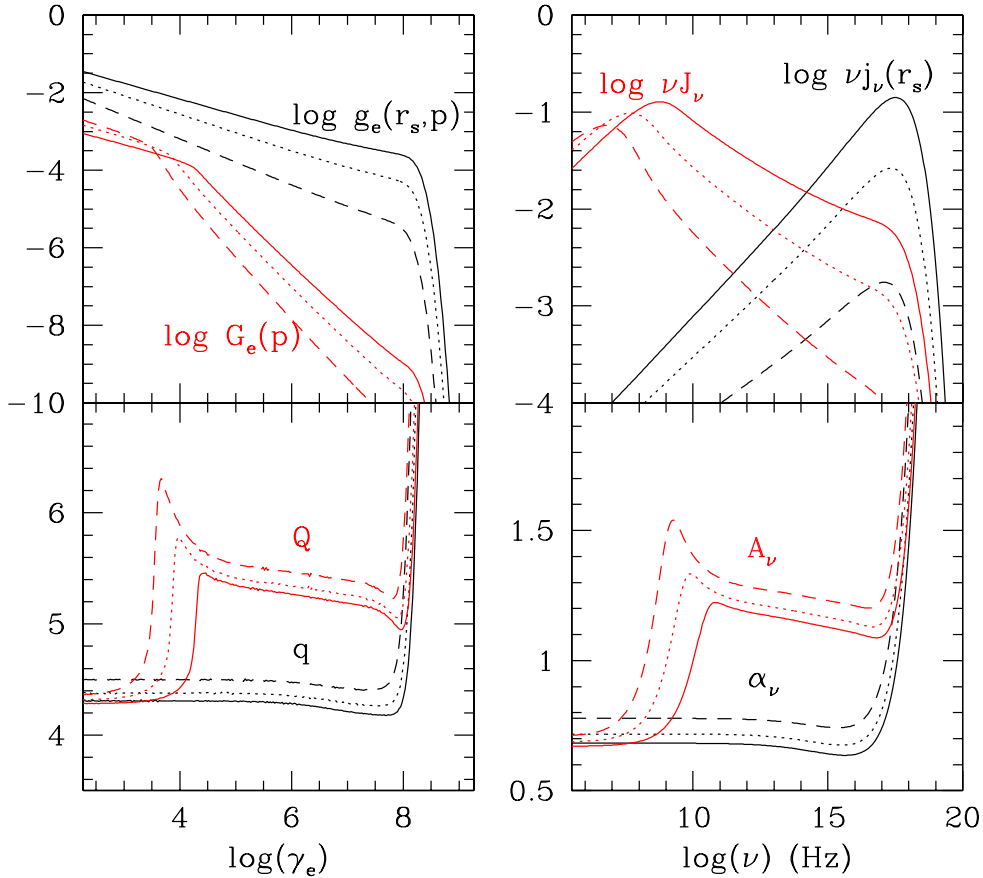}
\caption{ 
Same as Fig. 4 except that the results for 
the spherical shock shown in Fig. 3 are plotted for 
$t_{\rm age} = (1.7,~ 4.3,~ 8.7)\times 10^7 \yr$  (solid, dotted, and dashed lines, respectively).
}
\end{figure*}

\begin{figure*}[t1]
\centering
\includegraphics[trim=5mm 2mm 2mm 2mm, clip, width=135mm]{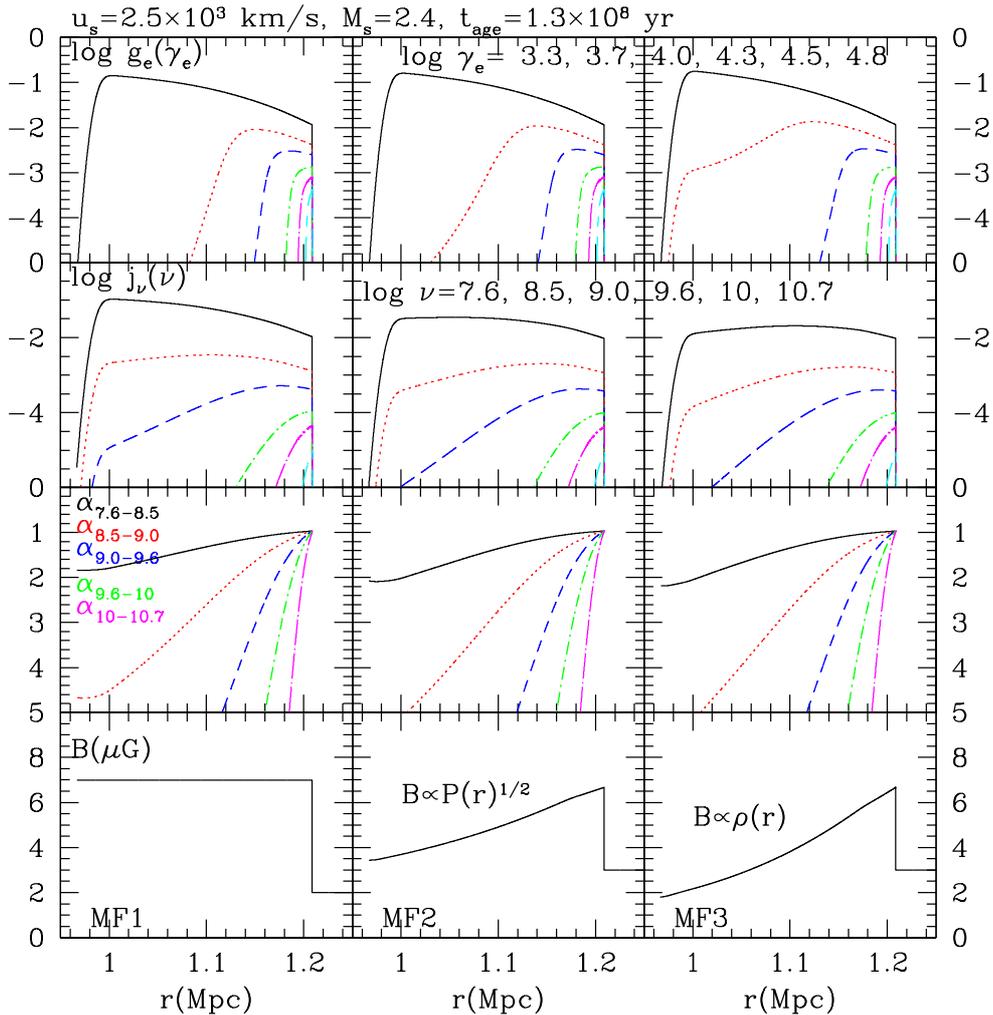}
\caption{ 
Spherical shock models with three different magnetic field profiles: {\bf MF1}, {\bf MF2}, and {\bf MF3}. 
The results are shown for
$t_{\rm age}=1.3\times 10^8 \yr$, when $u_s=2.5\times 10^3 \kms$ and $M_s=2.4$.
Spatial distributions of the electron distribution function, $g_e(r,\gamma_e)$, the synchrotron emissivity, 
$j_{\nu}(r,\nu)$, 
the spectral index, $\alpha_{\log \nu_1 - \log \nu_2}$, and
magnetic field strength, $B(r)$ (from top to bottom panels).
The line types for $g_e$, $j_{\nu}$, and $\alpha_{\log \nu_1 - \log \nu_2}$ are the same as in Figs. 2 and 3.
Note that $g_e$ and $j_{\nu}$ are plotted in arbitrary units.
}
\end{figure*}
\begin{figure*}[t!]
\centering
\includegraphics[trim=4mm 55mm 1mm 15mm, clip, width=135mm]{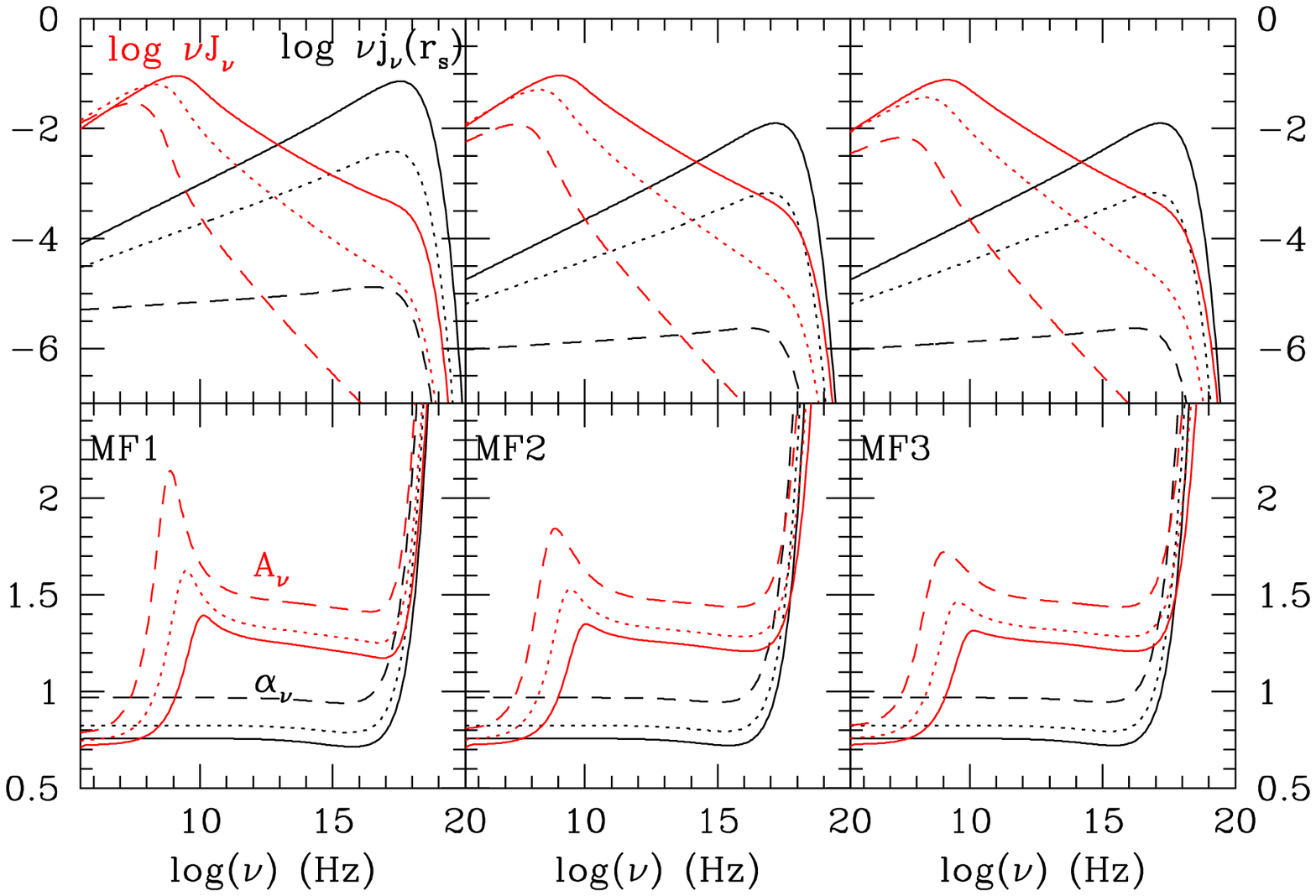}
\caption{ 
Spherical shock models with three different magnetic field profiles shown in Fig. 6.
Top: 
synchrotron spectral distribution at the shock position, $\nu j_{\nu}(r_s)$, 
and integrated over the simulation volume, $\nu J_{\nu}$. 
Bottom: synchrotron spectral index, $\alpha_{\nu}$ for $j_{\nu}(r_s)$ 
and $A_{\nu}$ for $J_{\nu}$.
The results are shown for $t_{\rm age}=(0.33, 0.67, 1.3)\times 10^8 \yr$ (solid, dotted, dashed lines, respectively).
Note that $j_{\nu}$ and $J_{\nu}$ are plotted in arbitrary units.
}
\end{figure*}

\section{DSA Simulation Results}

\subsection{Planar versus Spherical Shocks}

In this section we compare the electron acceleration at a spherical shock with
that at a planar shock with similar properties.
First, the planar shock is characterized by the following parameters:
the shock speed, $u_s=3.5\times 10^3\kms$, the sonic Mach number, $M_s=4.5$, 
the preshock density, $n_{H,1}=10^{-3} \cm3$, and the preshock and postshock
magnetic field strength, $B_1=2\muG$ (for $x>x_s$) and $B_2=7\muG$ (for $x<x_s$).

In the upper panel of Fig. 2, we show the spatial profiles of the electron
distribution function, $g_e(x,\gamma_e)$ for different energies, 
$\log \gamma_e=3.3,$ 3.7, 4.0, 4.3, 4.5, and 4.8.
Note that the shock location is shifted slightly from left to right at later time epochs,
$t_{\rm age}= (1.8, 4.4, 8.0)\times 10^7 \yr$.
The downstream width of $g_e$, $L(\gamma_e)$, 
for low energy electrons with $\log \gamma_e \lesssim  3.3$ increases
simply as $l_{\rm adv}= (u_s/\sigma) t_{\rm age} \approx 100{\rm kpc} (t_{\rm age}/10^8\yr)$.
For high energy electrons, on the other hand, $L(\gamma_e)$ decreases with $\gamma_e$,
since the radiative cooling time scales inversely with electron energy as in Eq. (\ref{trad}).
As can be seen in the figure, for $t_{\rm age}>4.4\times 10^7 \yr$, for instance, 
the width $L(\gamma_e)$ for $\gamma_e\ge 10^4$
becomes constant in time ($t_{\rm age}> t_{\rm rad}$)
as $L(\gamma_e)\approx (u_s/\sigma) t_{\rm rad}(\gamma_e)$ 
(see blue dashed, green dot-dashed, magenta dot-long dashed, and cyan dash-long dashed lines).

From Eq. (\ref{fpeak}), the electrons with $\gamma_e$ are expected to have the peak
of synchrotron emission at $\nu_{\rm peak} \approx 1{\rm GHz} (\gamma_e/10^4)^2$.
The middle panel of Fig. 2 shows the synchrotron emissivity, $j_{\nu}(x,\nu)$, for
the peak frequency that corresponds to the Lorentz factor chosen in the upper panel
(\ie $\log \nu= 7.6$, 8.5, 9.0, 9.6, 10., and 10.7), at the same epochs as in the upper panel.
Note that the spatial distribution of synchrotron emission at $\nu_{peak}$
is much broader than that of electrons with the corresponding $\gamma_e$.
This is because more abundant lower energy electrons also 
contribute significantly to the synchrotron emissivity at $\nu_{peak}$.
For example, the spatial profile of $j_{\nu}({\rm 1GHz})$ (blue dashed lines) 
is much broader and decreases much gradually, compared to that of $g_e$ of the corresponding 
$\gamma_e=10^4$ (also blue dashed lines) with $L(\gamma_e=10^4)\approx 40$~kpc.
So the downstream width of synchrotron emitting volume would be greater 
than that of the spatial distribution of $g_e(\gamma_e)$ for the value of $\gamma_e$ related by Eq. (\ref{fpeak}),
that is, $L(\nu_{peak}) > L(\gamma_e)$,
if the width is constrained by the cooling time $t_{\rm rad}$.
Note that $g_e$ and $j_{\nu}$ are plotted in arbitrary units in Fig. 2.

The bottom panel of Fig. 2 shows the spectral index which is defined as
\begin{equation}
\alpha_{\log \nu_1 - \log \nu_2}=- {{d\ln j_{\nu} }\over {d\ln \nu}}
\end{equation} estimated between $\nu_1$ and $\nu_2$. 
The peak frequencies chosen in the middle panel are adopted for $\nu_1$ and $\nu_2$, so
$\alpha_{7.6-8.5}$ (black solid line),
$\alpha_{8.5-9.0}$ (red dotted), 
$\alpha_{9.0-9.6}$ (blue dashed), 
$\alpha_{9.6-10}$ (green dot-dashed), and
$\alpha_{10-10.7}$ (magenta dot-long dashed) are plotted. 
The characteristic scale over which $\alpha_{\log \nu_1 - \log \nu_2}$ increases is governed 
mainly by the length scale of $j_{\nu}$ at the higher frequency $\nu_2$.
For example, the spectral index $\alpha_{9.0-9.6}$ (blue dashed line in the bottom panel) behaves similarly
to the emission $j_{\nu}$ at $\log \nu=9.6$ (green long dashed line in the middle panel).

For the spherical shock, we adopt a Sedov-Taylor similarity solution
with the shock speed, $u_{s,i}=u_o (t_i/t_o)^{-3/5}$, as the initial set up.
Here $t_i$ is the initial time from which electrons are injected at the shock
and $u_o=2.3\times 10^3\kms$ and $t_o=1.73\times 10^8\yr$.
For $t_i=0.5t_o$, the initial shock speed, $u_{s,i}\approx 3.5\times 10^3\kms$ and $M_s\approx 4.5$,
which are similar to the parameters for the planar shock shown in Fig. 2. 
Since the sonic Mach number is not large and the shock compression ratio is $\sigma=\rho_2/\rho_1\approx 3.5$,
the shock evolution behaves slightly differently from the true Sedov-Taylor solution. 
So at later time the shock speed evolves only approximately as $u_s(t)\approx u_o(t/t_o)^{-3/5}$.
Again note that only the structure and evolution of the flow velocity $u(r,t)$, is fed 
into the diffusion convection Eq. (\ref{diffcon}) and the other gasdynamical
variables do not influence the DSA process.
For the comparison with the planar shock, the magnetic field strengths are assumed to be constant in both space
and time with $B_1=2\muG$ and $B_2=7\muG$.
In the next section, we will consider several models with variable magnetic field strength,
$B(r,t)$ behind the shock.

The characteristics of the spherical shock model shown in Fig. 3 can be compared directly with 
those of the planar shock shown in Fig. 2.
Here the results are shown at the time elapsed from the initial injection, 
$t_{\rm age} \equiv (t-t_i)= (1.7, 4.3, 8.7)\times 10^7 \yr$, during which
the shock speed decreases from $3.5\times 10^3 \kms$ ($M_s=4.5$) to $1.5\times 10^3 \kms$ 
($M_s=3.1$).
The obvious feature, which is different from the planar shock case, is that
the injected particle flux decreases as the spherical shock slows down in time.
As a result, the amplitude of $g_e(\gamma_e)$ increases downstream away from the shock at a given time,
especially in the case of low energy electrons, which simply advect downstream without much
cooling.
So low frequency ($\nu\lesssim 360$~MHz) radio emission may increase towards downstream as well.
For high energy electrons, on the other hand,
the decline of $g_e(\gamma_e)$ due to cooling dominates over
the effect of decreasing injection flux.
Thus radio emission at higher frequencies ($\nu \gtrsim 1$~GHz) should decrease downstream behind the shock.
Again note that $g_e$ and $j_{\nu}$ are plotted in arbitrary units in Fig. 3.

Similar to the planar shock model, the spatial distribution of synchrotron emission at
$\nu_{peak}$ is much broader than that of $g_e(\gamma_e)$ for the corresponding $\gamma_e$.
For example, again the width of $g_e$, $L(\gamma_e=10^4)\approx 40$~kpc (blue dashed line in the top panel),
while the width of $j_{\nu}$, $L(\nu\approx 1{\rm GHz}) \approx 100$~kpc (blue dashed line in the middle panel).
But the spectral index $\alpha_{9.0-9.6}$ between 1~GHz and 4~GHz (blue dashed line in the bottom panel)
follows the spatial profile of the emission $j_{\nu}$ at 4~GHz (green dot-dashed line in the middle panel).
Thus the main difference between a planar shock and a decelerating spherical shock is
the decline of the injected particle flux in time, leading to possible increase of
low frequency radio emission downstream away from the shock.

Next we attempt to find any spectral signatures imprinted in the synchrotron emission spectrum
in the case of the spherical shock case.
In the left panels of Fig. 4, we show, at the three epochs, the electron spectrum at the shock, $g_e(x_s,p)$, 
and the volume integrated electron spectrum $G_e(p)=p^4 F_e(p)$, where $F_e(p)=\int f_e(p,x) dx$,
and their slopes, $q=-d \ln f_e(x_s)/d \ln p$, and
$Q=-d \ln F_e/d \ln p$. 
The slope from the simulation results is consistent with the test-particle slope $q=4.2$ for $\sigma=3.5$,
and the slope for the volume integrated spectrum, $Q=q+1$, above the break momentum $p_{\rm e,br}$.
Note that $\gamma_e = p/m_e c$ for relativistic energies.
As can be seen in Fig. 4, the electron energy spectrum at the shock has reached steady
state at these ages. The volume integrated spectrum shows the gradual decline of the break
Lorentz factor, $\gamma_{\rm e,br}$ with the shock age.

The same kinds of quantities for the spherical shock are shown in the left panels of Fig. 5.
We can see the time evolutions of the electron spectrum as the shock slows down in time.
Since the shock Mach number decreases, the slope $q$ for $g_e(r_s,p)$ increases in time,
leading to a nonlinear change of the volume integrated spectrum, $G_e(p)$.
So the slope $Q$ no longer follows the simple $q+1$ steepening above $p_{\rm e,br}$.
In addition to usual radiative cooling, 
both the reduced injected particle flux and steepening of $g_e(r_s,p)$ due to the slowing-down and 
weakening of the shock affect the spectral shape of $G_e(p)$.

In addition, we show the synchrotron emission spectrum at the shock,
$j_{\nu}(x_s)$ in Fig. 4 and $j_{\nu}(r_s)$ in Fig. 5, 
and the volume integrated emission spectrum, $J_{\nu}=\int j_{\nu}(x) dx$ in Fig. 4 and
$J_{\nu}=\int j_{\nu}(r) 4\pi r^2 dr$ in Fig. 5.
Their respective spectral indexes, $\alpha_{\nu}=- {d\ln j_{\nu} }/ {d\ln \nu}$ and
$A_{\nu}=- {d\ln J_{\nu} }/ {d\ln \nu}$ are shown in the lower right panels of Figs. 4 and 5.
For the planar shock with $\sigma=3.5$, 
the test-particle spectral indexes are $\alpha_{\nu}=0.6$ and $A_{\nu} = 1.1$ for $\nu>\nu_{\rm br}$,
as can be seen in Fig. 4.
We note that the transition of the volume-integrated spectral index from $A_{\nu}=\alpha_{\nu}$ 
to $A_{\nu}=\alpha_{\nu}+0.5$ is more gradual,
compared to the transition of the particle slope from $Q=q$ to $Q=q+1$.
For the shock parameters chosen here, at the shock age of $\sim 10^8 \yr$ this transition occurs over 
the broad frequency range, $0.01~{\rm GHz}\lesssim \nu \lesssim 10~{\rm GHz}$.

For the spherical shock shown in Fig. 5, on the other hand, the volume integrated spectrum, $J_{\nu}$, 
develops a concave curvature
and its spectral index $A_{\nu}$ deviates from the simple relation of $A_{\nu}=\alpha_{\nu} + 0.5$. 
Again, at the shock age of $\sim 10^8 \yr$ the nonlinear curvature occurs over 
the broad frequency range, $0.01~{\rm GHz}\lesssim \nu \lesssim 10~{\rm GHz}$.
Thus the simple relation, $A_{\rm integ}\approx \alpha_{\rm inj} + 0.5$,
which is often invoked for the DSA origin of synchrotron emission spectra, 
should be applied with some caution in the case of evolving shocks.
Again in Figs. 4 and 5, $g_e$, $G_e$, $j_{\nu}$ and $J_{\nu}$ all are plotted in arbitrary units.

\subsection{Spherical Shocks with Different Magnetic Field Models}

Self-excitation of MHD waves and amplification of magnetic fields via plasma instabilities
are the integral parts of DSA at {\it strong} collisionless shocks.
CRs streaming upstream in the shock
precursor excite resonant Alfv\'en waves with wavelengths 
comparable with the CR proton gyroradii, and turbulent magnetic
fields can be amplified into the nonlinear regime \citep{bell78,lucek00}. 
In addition, the
nonresonant fast-growing instability driven by the CR current can amplify the magnetic field by orders
of magnitude \citep{bell04,riqu09}. 
It was shown that, in the case of strong shocks ($M_s>10$), 
magnetic amplification (MFA) via streaming instabilities and Alfv\'enic drift can
lead to significant nonlinear feedback on the shock structure and 
the predicted CR spectrum \citep[e.g.,][]{capri12,kangJ12}.

In a recent study of hybrid plasma simulations, \citet{capri14b} has shown that the MFA
factor increases with the Alfvenic Mach number as $\langle \delta B/B\rangle^2 \sim$ $3 M_A (P_{cr,2}/\rho_1 u_s^2)$,
where $\delta B$ is the turbulent magnetic fields perpendicular to the mean background magnetic fields.
For the sonic Mach number in the range of $3\lesssim M_s \lesssim 5$, the CR proton acceleration efficiency varies as
$P_{cr,2}/\rho_1 u_s^2 \sim 0.01 -0.05$ \citep{kang13,capri14a}.
For typical cluster shocks with $M_A\sim 10$, 
the MFA factor due to the streaming stabilities is expected be rather small, 
$\langle \delta B/B\rangle^2 \sim 0.3-1.5$.
How magnetic fields may be amplified at weak shocks in high beta ICM plasmas
has not yet been fully understood.
As in the previous study of \citet{kang12}, we adopt the postshock magnetic field strength
$B_2\sim 7\muG$ in order to produce the observed width of the Sausage relic ($\Delta l \sim 50$~kpc),
while the preshock magnetic field strength is chosen to be $B_1\sim 2-3 \muG$.
Note that $B_1$ cannot be constrained directly from observations, since we do not detect any
radio emission upstream of radio relic shocks. 
Thus $B_1$ can be adjusted so that $B_2$ becomes about 7 $\muG$ after considering MFA via plasma instabilities or
the compression of the perpendicular components of magnetic fields across the shock.
In any case, the postshock electron energy spectrum and the synchrotron radiation spectrum do not
depend sensitively on the value of $B_1$ as long as it is a few microgauss.

Here we consider the following three magnetic field models:

{\bf MF1}: $B_1=2\muG$ \& $B_2=7\muG$.

{\bf MF2}: $B_1=3\muG$, $B_2=B_1 \sqrt{1/3+2\sigma^2/3} $, 

~~~~~~~~\& $B(r)= B_2 \cdot (P(r)/P_2)^{1/2}$ for $r<r_s$.

{\bf MF3}: $B_1=3\muG$, $B_2=B_1 \sqrt{1/3+2\sigma^2/3}$, 

~~~~~~~~\& $B(r)= B_2 \cdot (\rho(r)/\rho_2)$ for $r<r_s$.

In {\bf MF2} and {\bf MF3} models, the postshock magnetic field energy is assumed to be dissipated,
and so $B(r)$ decreases with the gas pressure or density behind the shock.
As shown in the bottom panels of Fig. 6, the decay length scale is about 100-150~kpc 
in {\bf MF2} and {\bf MF3} models.

As in the previous section, we consider a spherical blast-wave, $u_{s,i} \approx u_o(t_i/t_o)^{-3/5}$, 
where $u_o=3.0\times 10^3\kms$ and $t_o=1.3\times 10^8\yr$
(slightly faster than the shock adopted in Fig. 3).
The calculations were started at $t_i=0.5t_o$ with $u_{s,i}\approx 4.5\times 10^3\kms$ and $M_s\approx 4.3$
and ended at $t_f=1.5 t_o$ with $u_{s,f}\approx 2.5\times 10^3\kms$ and $M_s\approx 2.4$.
The results of these three models at the final time ($t_{\rm age}=t-t_i \approx 1.3\times 10^8 \yr$)
are compared in Fig. 6.
As in the spherical shock model shown in Fig. 3, 
the spatial profile of $g_e$ for low energy electrons increases downstream
away from the shock. 
The distributions of $g_e(\gamma_e\approx 5\times 10^3)$ (red dotted lines) show some differences among
the three models due to different synchrotron cooling rates.
The spatial profiles of $j_{\nu}$ at low frequencies 
(e.g., 360~MHz in red dotted lines) show rather weak dependences on the $B(r)$ model, 
while those at higher frequencies (e.g., 1~GHz in blue dashed lines) behave similarly in all three cases.
Moreover, the steepening patterns of $\alpha_{\nu}$ behind the shock show no significant variations 
among the three models.

Thus it may be difficult to differentiate the postshock magnetic field profile from the
spatial variation of spatially resolved radio flux, $S_{\nu}$, or spectral index, $\alpha_{\nu}$.
In other words, signatures imprinted on synchrotron emission due to different $B(r)$ behind 
the shock would be too subtle to detect.
There are two reasons that may explain this prediction.
Because the iC cooling with $B_{\rm rad}=4.7\muG$ provides the baseline for the cooling rate, $b(p)$,
the difference in the synchrotron cooling due to different $B(r)$ has relatively minor effects.
Also any sharp features in the distribution of $f_e(r,p)$ would be smeared out
in the distribution $j_{\nu}(r)$, since the synchrotron emission at a given frequency comes
from electrons with a somewhat broad range of $\gamma_e$.

Finally, in Fig. 7 we compare the synchrotron radiation spectrum at the shock, $j_{\nu}(r_s)$, and the volume integrated 
radiation spectrum, $J_{\nu}$, for the three magnetic field models at $t_{\rm age}=(0.33, 0.67, 1.3)\times 10^8 \yr$.
Departures from the predictions for the test-particle planar shock are most severe in {\bf MF1} model,
while it becomes relatively milder in the two models with decaying $B(r)$ behind the shock.
As expected, any variations in the spatial profiles of $f_e(r,p)$ and $B(r)$ are averaged out in
the volume integrated quantities such as $F_e(p)$ and $J_{\nu}$. 

In conclusion, any features in the spectral curvature, $dA_{\nu}/d\nu$, or the steepening of spectral index, $d\alpha_{\nu} /dr$ in the postshock region,
should be affected in principle by both the time-dependent evolution of
the shock and the spatial variation of postshock magnetic fields.
But the effect of decaying magnetic fields may be too subtle to detect due to the dominance of
iC scattering off the CBR.

\section{Summary}

We have performed time-dependent DSA simulations for cosmic-ray (CR) electrons 
at a planar shock with a constant speed 
and decelerating spherical shocks similar to Sedov-Taylor blast waves.
These shocks were chosen to have parameters relevant for weak shocks 
typically found in the outskirts of galaxy clusters:
$u_s\approx (1.5-4.5)\times 10^3 \kms$, $M_s\approx 2.5-4.5$, $B_1\approx 2-3 \muG$, and $B_2\approx 7 \muG$.
For spherical shocks, we used a one-dimensional spherical version of the CRASH code, in which
a co-expanding spherical grid was adopted in order to optimize the shock tracking
and adaptive mesh refinement features \citep{kj06}.
The diffusion convection equation for the pitch-angle-averaged phase space distribution function 
for CR electrons in the test-particle regime was solved
along with the usual gasdynamic conservation equations
without including the dynamical feedback of the CR proton pressure.
The synchrotron emission from CR electrons accelerated at these model shocks was estimated by
adopting several models for magnetic field profile, $B(r)$, and the electron energy spectrum,
$N_e(r,\gamma_e)$ from the DSA simulations.

Since the time scales for electron acceleration and cooling are much shorter than
the dynamical time scale, the electron energy spectrum at the shock, $f_e(r_s,p,t)$, has reached 
the steady state in the planar shock model, as can be seen in Fig. 4.
The slope of the volume-integrated momentum distribution, $F_e(p)$, is simply $Q=q+1$ above
the break momentum ($p>p_{\rm e,br}$). 
However, the spectral index of the volume-integrated synchrotron emission spectrum, $J_{\nu}$, 
steepens gradually to $A_{\nu}=\alpha_{\rm inj}+0.5$
over the broad frequency range, $\sim (0.1-10)\nu_{\rm br}$.
They are consistent with the predictions of the DSA theory in the test-particle limit.

In the spherical shock cases, the electron energy spectrum at the shock has reached 
the steady state defined by the {\it instantaneous} shock parameters (\ie $u_s(t)$, $r_s(t)$ and $M_s(t)$),
as shown in Fig. 5.
So the synchrotron radiation spectra at the shock location could be described properly 
by the test-particle DSA predictions. 
However, the volume integrated spectra of $F_e(p,t)$ and $J_{\nu}(\nu,t)$ both evolve differently 
from those of the planar shock, depending on the time-dependent evolution of the shock parameters.
As a result, the slopes, $Q$ and $A_{\nu}$, exhibit some nonlinear signatures that come from 
decreasing particle flux and declining shock Mach number (see Fig. 5).
This suggests that, in the case of evolving shocks,
one needs to be cautious about interpreting observed radio spectra
by adopting simple DSA models in the test-particle regime. 

In addition, we have considered spherical shock models with decaying postshock magnetic fields,
in which $B(r)$ decreases behind the shock with a scale height of 100-150~kpc, as shown in Fig. 6.
Because inverse-Compton scattering off the cosmic background photons with $B_{\rm rad}\approx 4.7\muG$
(for $z=0.2$ adopted here as a reference epoch) provides the baseline cooling rate, the electron energy 
spectra do not depend sensitively on the $B(r)$ profile.
We also find that the impacts of the different $B(r)$ profile 
on the spatial profile of synchrotron emission, $j_{\nu}(r)$, and its spectral index, $\alpha_{\nu}(r)$,
is rather minimal, since electrons in a somewhat broad range of $\gamma_e$ contribute
emission to $j_{\nu}$ at a given frequency.
Of course, any nonlinear features due to the spatial variations in $f_e(r,p)$ and $B(r)$ are mostly
averaged out and may leave only subtle signatures in the volume integrated spectrum, $J_{\nu}$, and its spectral index, $A_{\nu}$.

\acknowledgments

This work was supported by a 2-Year Research Grant
of Pusan National University.


\end{document}